
\magnification=\magstep1
\overfullrule=0pt
\def\*{^}
\def\h{h\*\vee} \def\c{\chi} \def\C{\chi\* *}
\def\Q{{\bf Q}}
\def\L{\Lambda}   \def\la{\lambda}
 \def\Z{{\bf Z}}
\def\u{\tau}       \def\g{{\hat g}}
\def\equi{\,{\buildrel \rm def \over =}\,}
\def\sp{\;}    \def\eg{{\it e.g.}$\sp$} \def\ie{{\it i.e.}$\sp$}

\def\i{\item}
\def\sp{\,\,\,}    \font\huge=cmr10 scaled \magstep2
\def\d{\delta}
\def\LSW{[20]} \def\CA{[1]} \def\GW{[2]} \def\KP{[4]} \def\VE{[13]}
\def\FKSV{[3]} \def\CIZ{[5]} \def\MS{[6]} \def\SY{[7]} \def\KA{[8]}
\def\Kas{[9]} \def\BER{[26]} \def\RT{[15]} \def\WA{[14]} \def\HET{[16]}
\def\COMM{[12]} \def\CR{[27]} \def\CS{[19]} \def\ITZ{[11]} \def\KAC{[17]}
\def\KMPS{[18]} \def\RTW{[23]} \def\RTWb{[21]} \def\GA{[10]}\def\RUE{[24]}
\def\ALZ{[25]} \def\BI{[22]} \def\ITCO{[11,12]} \def\CIK{[5,8]}
\def\VEFK{[13,3]} \def\KK{[17,18]} \def\RTWA{[14,15]} \def\LAT{[19,20]}
\def\Gal{[28]}\def\Csgl{[19,28]}

{ \nopagenumbers
\rightline{April, 1993}
\rightline{final draft}
\bigskip \bigskip
\centerline{{\bf \huge The Low Level Modular Invariant Partition Functions}}
\bigskip\centerline{{\bf \huge of Rank-Two Algebras}}
\bigskip \bigskip
\centerline{Terry Gannon}
\centerline{{\it Mathematics Department, Carleton University}}
\centerline{{\it Ottawa, Ont., Canada K1S 5B6}}\bigskip
\centerline{Q. Ho-Kim}
\centerline{{\it Physics Department, Laval University}}
\centerline{{\it Sainte-Foy, Que., Canada G1K 7P4}}\bigskip
\bigskip \bigskip \centerline{{\bf Abstract}}\bigskip

Using the self-dual lattice method, we make a systematic search for modular
invariant partition functions of the affine algebras $g\*{(1)}$ of $g=A_2$,
$A_1+A_1$, $G_2$, and $C_2$. Unlike previous computer searches, this method is
necessarily complete. We succeed in finding all physical invariants for $A_2$
at levels $\le 32$, for $G_2$ at levels $\le 31$, for $C_2$ at levels $\le
26$, and for $A_1+A_1$ at levels $k_1=k_2\le 21$.  This work thus completes a
recent $A_2$ classification proof, where the levels $k=3,5,6,9,12,15,21$ had
been left out.  We also compute the dimension of the (Weyl-folded) commutant
for these algebras and levels.
\vfill \eject }
\pageno=1
\bigskip \centerline{{\bf I. INTRODUCTION}} \bigskip

     The subject of this paper is the classification of the partition functions
of some conformal field theories. The problem can be stated in general terms as
follows.  Consider a conformal field theory that has an operator algebra
decomposable into a pair of commuting holomorphic and antiholomorphic chiral
algebras, ${\g_L}$ and ${\g_R}$, and a Hilbert space which can be written as a
finite sum of irreducible representations $(\la_L,\la_R)$ of
${\g_L\times\g_R}$ with multiplicity $ N_{\la_L \la_R}$.  With each such
representation $(\la_L,\la_R)$, associate a pair of characters $\c_{\la _L}$
and $\c_{\la_R}$.  Then define the partition functions of the theory as
combinations of bi-products of characters of the form

   $$Z=\sum N_{\la_L \la_R} \c_{ \la_L}\,\C_{\la_R}. \eqno(1.1)$$

The problem is to find, for a given algebra $(\g_L, \g_R)$, all combinations
(1.1) such that: (P1) $Z$ is invariant under transformations of the modular
group; (P2) the coefficients $N_{ \la_L \la_R}$ are non-negative integers; and
(P3) $N_{11}=1$ if $\la=1$ denotes the vacuum.

Any function $Z$ satisfying the modular invariance condition (P1)  will be
called an {\it invariant\/}; if in addition it satisfies the condition that
each coefficient $N_{\la_L\la_R}\geq 0$, then it is said to be a {\it positive
invariant\/}; and finally, if the conditions (P1), (P2) and (P3) are all
met, then it is considered to be a {\it physical invariant\/}.  Clearly
any conformal theory must have {\it at least\/} these three properties to be
physically meaningful.

Such functions appear in the context of statistical physics, for example
in the early work of Cardy \CA{} when he formulated a conformal system on a
finite rectangular strip with periodic boundary conditions. They also arise in
the theory of strings, for example in the classical paper by Gepner and Witten
\GW{} where the authors discussed the compactification of closed strings on Lie
group manifolds.  The action for a closed string is the same as that for a
nonlinear sigma model with properly chosen couplings (\ie the
Wess-Zumino-Witten model) and is known to be not only conformally invariant but
also invariant under the much larger current algebra induced by the underlying
group, which forms itself a subclass of the Kac-Moody algebras.  In the
context of the closed string theory the object of interest is the factor
pertaining to the group manifold of the vacuum-to-vacuum string amplitude in
the lowest-order (one-loop) approximation.

It is quite possible that, among all such invariants, some particular
ones may lead to phenomenologically interesting, even viable, string models, as
some recent analyses \FKSV{} have suggested.  Also, it has long been a hope by
many that a good understanding of the partition functions could make a positive
contribution to the broader and as yet incomplete task of classifying conformal
field theories. This hope is justified partly by the fact that a conformal
field theory can be identified in some sense through its partition functions,
and partly by some recent progress in the computing and understanding of the
invariants.  The characters of an algebra, considered in their full dependence
on the coordinates of the space on which they are defined, are linearly
independent \KP. And so are functions of the type $\c_{\la_L}\C_{\la_R}$. It
means that two theories with the same partition functions are identical and,
conversely,  different theories must correspond to different partition
functions.

Many modular invariant partition functions have already been known, and much
insight has been gained through the efforts of many, in particular \CIZ, \MS,
and \SY, among others.  But classification proofs, which determine all
the physical invariants that belong to a certain class, exist only for a few
cases, namely, the untwisted Kac-Moody algebra $A_1\*{(1)}$ \CIK{} and the
coset models based on it, such as the minimal unitary Virasoro models \CIZ{}
and the  $N=1$  minimal superconformal models \Kas; recently a classification
proof for $A_2\*{(1)}$ has also been obtained \GA.  These proofs hold for an
arbitrary {\it level\/} (a number through which affine representations can be
related to finite ones). In addition, the complete list of level-one
partition partitions for the simple Lie algebras $A_n\*{(1)}$, $B_n\*{(1)}$,
$C_n\*{(1)}$, $D_n\*{(1)}$, and the five exceptional algebras has been found
\ITCO.

Systematic numerical searches for invariants of algebras are useful in guiding
and confirming theoretical analyses, and in discovering new exceptional
invariants. Several of these searches have already been made. For example,
\VEFK{} use a procedure based on Verlinde's formula rephrased as an
eigenvalue problem.  However in general these searches are not necessarily
complete -- \ie it is possible that an elusive and presently unknown
exceptional invariant could escape them (\eg \VEFK{} place an upper
bound on the sizes of the coefficients $N_{\la_L\la_R}$). In the work reported
in the present paper, we follow yet another method, proposed by Warner \WA{}
and independently by Roberts and Terao \RT.  In their approach, invariants are
generated by using the Weyl-Kac formula of characters and the theta functions
associated with even self-dual lattices.  It has been shown \COMM{} that this
method is complete: it will yield all the invariants of a given class. It is a
number-theoretic technique that treats all invariants -- exceptional invariants
as well as members of infinite series -- on the same footing.  It is quite
practical, at least for low rank algebras and small levels, as in the cases we
considered.

Its completeness and practicality make this lattice method ideally suited
to finish off the $A_2\*{(1)}$ classification. In \GA{} this classification
was accomplished for all levels $k$, except for $k=3,5,6,9,12,15,21$. These
levels survived because the proof in \GA{} broke down for them. In this
paper our analysis of $A_2\*{(1)}$ includes those remaining levels, thus
completing the classification for $A_2\*{(1)}$.

We will implicitly assume throughout this paper that the algebras $\g_L,\g_R$
of the holomorphic and antiholomorphic sectors of the theory are identical.
This is the situation most commonly considered; the problem of asymmetric
(heterotic) invariants is addressed in \HET.

In Sec.~II we will describe the method. In Sec.~III we discuss the calculations
of the invariants for the affinizations of the four rank-two algebras: $A_2$
at levels $\le 32$; $G_2$ at levels $\le 31$; $C_2$ at levels $\le 26$; and
$A_1+A_1$ at levels $k_1=k_2\le 21$. The results of our calculations, namely
the complete set of all physical invariants for those algebras and levels,
are given in Sec.~IV. Our concluding remarks are found in Sec.~V. The Appendix
describes in more detail how we find all of the desired lattices.

\bigskip \bigskip
		 \centerline{{\bf II. THE LATTICE APPROACH }}
\bigskip\nobreak

In this section we will briefly review some basic elements of the affine
Kac-Moody algebras \KK{} and describe the lattice method of
Roberts-Terao-Warner \RTWA{} that we used in our numerical calculations.  We
will consider here simple algebras $g$, but analogous statements hold also for
$g$ semi-simple.

   Let $g$ be a  finite-dimensional simple Lie algebra of rank $r$, and let
$\alpha_1\*{\vee},\ldots,\alpha_r\*{\vee}$ be its simple coroots.  The
fundamental weights of $g$ are the vectors $\beta_1,\ldots,\beta_r$ defined by
the inner products $\beta_i\cdot\alpha_j\*\vee=\delta_{ij}$. Let $\rho$ be the
sum of all the fundamental weights, $\rho=\sum \beta_i$.  The coroots span the
coroot space. By {\it coroot lattice} of $g$, denoted by $M=M_g$, we mean the
integral span of its coroots, a subspace of the coroot space. The dual $M\* *$
of the coroot lattice is called the weight lattice, the set of integral weight
vectors. Any vector $\la\in M\* *$ can be decomposed as $\la=\sum_{i=1}\* r
m_i\beta_i$ where $m_i\in \Z$. If $\la$ is a highest weight, all
$m_1,\ldots,m_r$ are nonnegative integers.

     Let $\g=g\*{(1)}$ be the untwisted affine extension of $g$ defined by
a central element and a derivation.  Weight vectors of $\g$ may
then be denoted $\hat \la=(\la,k,n)$, where $\la$ is a weight of $g$, $n$ an
eigenvalue of the derivation, and $k$ an eigenvalue of the central element
(called the {\it level\/}).  Each unitary highest weight representation of $\g$
is associated with an extended Dynkin diagram with $r+1$ nodes labelled by
$m_0,m_1,\ldots,m_r$, where $m_1,\ldots,m_r$ are related to a horizontal
highest weight vector $\la$ in the usual way, and the label of the extra node
is given by $m_0=k-\sum_{i=1}\*{r}m_ia_i\*\vee$, where $k$ is the level of the
representation and $a_i\*\vee$ are the positive integral colabels of $g$.  The
number $1+\sum a_i\*\vee$ is known as the {\it dual Coxeter number} and denoted
by $\h$.

	An integrable highest weight representation of $\g$ is defined as one
with $m_0\ge 0$.  From this condition it follows that the level $k$ is also a
nonnegative integer since for an irreducible highest weight representation of a
simple Lie algebra $m_1,\ldots,m_r$ are nonnegative integers.  It is thus
convenient to introduce for each level $k=0,1,2,\ldots\,$  a subset of $M\* *$
containing the horizontal highest weight, $\la$, of each level-$k$ integrable
highest weight representation of $\g$:

$$ P_{+}(g,k)\equi\bigl\{\sum_{i=1}\* r m_i \beta_i\,|\, m_i\in \Z,\,\,
\,0\le m_i,\,\,\, \sum_{i=1}\* r m_i a_i\*\vee\le k\bigr\}.$$

\noindent For a given level, the number of horizontal highest weights is
finite.  In the context of a  conformally invariant current algebra theory,
each $\la\in P_{+}(g,k)$ is associated with an `integrable' primary field.
Only integrable fields are physically relevant; all of the other possible
primary fields are nonintegrable and, as such, irrelevant in the sense that any
correlation function containing one or more of these fields would vanish
identically. We will usually find it more convenient to use $\la+\rho$ in
place of the horizontal highest weight $\la\in P_+(g,k)$.

	The Weyl-Kac formula, which proves a very powerful tool in studying
modular invariance, expresses the character of an irreducible integrable
highest weight representation of $\g$ as a certain sum over the Weyl group of
$\g$. Since this Weyl group turns out to be a semidirect product of the finite
dimensional Weyl group, $W(g)$, and a lattice translation group, each summand
can be recast in terms of a {\it theta series} which takes care of the
translation group, leaving only the summation over $W(g)$ explicit. The result
is the celebrated Kac-Peterson formula \KP.

	Generally, given any lattice $\L$, the translate $v+\L$ of any vector
$v\in\Q\otimes \L$ defines a {\it glue class}.  The theta series of that glue
class is given by
$$\Theta\bigl(v+\L\bigr)(z|\u)\equi \sum_{x\in v+\L}\exp[
	 \pi i \u\,x\*2 +2\pi iz\cdot x].\eqno(2.1)$$
Here $\u\in {\bf C}$, and $z$ is a complex vector lying in ${\bf C}\otimes \L$.
If the lattice $\L$ is chosen to be positive definite this series converges,
and in fact is analytic for all such $z$ and any $\u$ in the upper half-plane.
Given any Euclidean lattice $\L$ and a positive number $\ell$, we write
$\L\!\*{(\ell)}$ for the positive definite scaled lattice $\sqrt{\ell}\,\L$ and
$\L\!\*{(-\ell)}$ for the corresponding negative definite scaled lattice.

	The Kac-Peterson character formula for any $\la\in P_{+}(g,k)$ can now
be written as: $$\eqalignno{\c_{\la+\rho}\*{g,k}(z,\u)=&{\sum_{w\in W(g)}
\epsilon(w) \,
\Theta\bigl({\lambda+\rho\over\sqrt{K}}+M\*{(K)}\bigr)(\sqrt{K}\,w(z)|\u) \over
D_g(z|\u)},&(2.2a)\cr D_g(z|\u)\equi &\sum_{w\in W(g)}\epsilon(w)\,
\Theta\bigl( {\rho\over\sqrt{\h}}+M\*{(\h)}\bigr)(\sqrt{\h}\,w(z)|\u),
&(2.2b)\cr}$$

\noindent  where $K=k+\h$ is the height of the representation, and
$\epsilon(w)=$det($w)\in \{\pm 1\}$ the ``parity'' of the transformation $w\in
W(g)$.

	Actually, the character should have been written as
$\c_{\la+\rho}\*{g,k}(u,z,\u)$ in its full dependence on all three coordinates
$z\in{\bf C}\otimes M$ and $u,\,\u\in {\bf C} $, Im$(\u)>0$. However, the
$u$-dependence is trivial, being an overall exponential, and will not play any
role in our study; for this reason and without loss in generality we set $u=0$.
On the other hand, it is important to maintain the full $z$-dependence;
otherwise the characters would cease to be linearly independent, and the
partition functions of two representations that are complex conjugates of each
other would become identical.

	So far $\c_\lambda\*{g,k}$ has been defined only for $\la-\rho\in
P_{+}(g,k)$.  However, for later purposes, it is convenient to use (2.2$a$)
to define $\c_\la\*{g,k}$ for {\it any\/} $\lambda\in M\* *$.  The following
observation \KAC{} permits us to evaluate these $\c_\la\*{g,k}$. For any
$\la\in M\* *$ and $k\geq 0$, there are two possibilities: either the character
vanishes
	  $$\c_{\la}\*{g,k}(z,\u)=0\eqno(2.3a)$$
for all $z$ and $\u$, or there exists a unique $w'\in W(g)$ and unique $\la'\in
P_{+}(g,k)$ such that $\la'+\rho=w'(\la)$ (mod $M\*{(K\* 2)}$) and
$$\c_{\la}\*{g,k}(z,\u)
	 =\epsilon(w')\,\,\c_{\la'+\rho}\*{g,k}(z,\u)\eqno(2.3b)$$
for all $z$ and $\u$.  We call $\tilde\epsilon(\la)$ the ``parity'' of $\la$
defined as $\tilde\epsilon(\la)=0$ if (2.3$a$) holds, and
$\tilde\epsilon(\la)=\epsilon(w')$ if (2.3$b$) does. In the latter case $[\la]$
will denote the unique $\la'+\rho=w'(\la)$ (mod $M\*{(K\* 2)}$).

	Let us now recall a few relevant facts about lattices \LAT.
An {\it integral\/} lattice is one in which all vectors give integral
dot products.  An {\it even\/} lattice is an integral lattice in which all
vectors have even norms.  An {\it odd\/} lattice is an integral lattice with at
least one odd-norm vector. A {\it self-dual\/} lattice $\L$ is one which equals
its dual $\L\* *$.  $\L$ is self-dual {\it iff\/} it is integral and has
determinant $|\L|=1$.  One calls a {\it gluing\/} $\L$ of some lattice $\L_0$
any lattice that can be written as a finite disjoint union of glue classes
of $\L_0$.  In other words, $\L_0$ is a sublattice in $\L$ of finite index, \ie
such that $\L/\L_0$ is a finite group.

	In the approach we are following, extensive use is made of Lorentzian,
or indefinite, lattices.  In such a lattice, every vector $x$ may be written
$(x_L;\,x_R)$, and the inner product of any pair of lattice vectors $x$ and
$x'$ is given by $x\cdot x'=x_L\cdot x_L'-x_R\cdot x_R'$.
In particular, let us consider the $2r$-dimensional indefinite lattice
$M_g\*{(K)}\oplus M_g\*{(-K)}$ that we shall call $\L\!\*{g,k}\equi
(M_g\*{(K)};\, M_g\*{(K)})$, with the height of $\g$, $K=k+\h$, being used
as the scaling factor.
Its elements are written as  $x=(x_L;\,x_R)$, where
$x_L,\,x_R\in M_g\*{(K)}$.  A gluing $\L$ of $\L\*{g,k}$ will look like
$\L=\L\!\*{g,k}[x_1,\ldots,x_n]$ $\equi \{ x+\sum_{i=1}\*n x_i l_i \,|\,
l_i\in\Z,\sp x\in \L\!\*{g,k}\}$, where the vectors
$x_1,\ldots,x_n\in\L\!\*{g,k*}$ are called the {\it glues} of the lattice
$\L$.  For example, take $x_i=(\beta_i/\sqrt{K}\,;\,\beta_i/\sqrt{K})$
where, as usual, $\beta_i$ are the fundamental weights of $g$. Then the gluing
$\L_D=\L\!\*{g,k} [x_1,\ldots,x_r]$, called the {\it diagonal gluing\/}, can be
seen to be both even and self-dual. (In the appendix we give a practical
test for determining whether or not a given gluing is self-dual.)

For any gluing $\L$ of the lattice $\L\!\*{g,k}$, let us define its level $k$
partition function as
 $$\eqalignno{Z_\L(g,k)(z_L z_R|\u) &\equi
 \sum_{w_L,w_R \in W(g)} \epsilon(w_L) \epsilon(w_R)\sum_{(x_L;x_R)\in \L}
 \exp[\pi i\, (\u x_L\*2- \u\* * x_R\*2)] &(2.4)\cr
 & \times \exp[2\pi i\sqrt{K}\, (\,w_L(z_L)\cdot x_L-w_R(z_R)\* *\cdot x_R\,)]
    /D_g(z_L|\u)D_g(z_R|\u)\* *.&\cr}$$

\noindent Because
$\L$ is a gluing of $\L\!\*{g,k}$, (2.2$a$) tells us we can
write $Z_\L(g,k)(z_Lz_R|\u)$ as a sum of terms proportional
to
    $$\c_{\la_L}\*{g,k}(z_L,\u)\,\c_{\la_R}\*{g,k}(z_R,\u)\* *\eqno(2.5)$$
for
$\la_L,\la_R\in M\* *$.
Eqs.(2.3$a$, $b$) further  tell us that $Z_\L(g,k)(z_Lz_R|\u)$ can be
written as a linear combination (over $\Z$) of terms of the form (2.5),
but with $\la_L-\rho$ and $\la_R-\rho$ now lying in $P_{+}(g,k)$, that is,

$$Z_\L(g,k)(z_Lz_R|\u)\,\equi\sum_{\la_L,\la_R\in P_++\rho}(N_\L)_{\la_L,\la_R}
   \; \c_{\la_L}\*{g,k}(z_L,\u)\,\c_{\la_R}\*{g,k}(z_R,\u)\* *,\eqno(2.6)$$

\noindent where $(N_\L)_{\la_L,\la_R}$ are integral scalars. For example,
the diagonal gluing $\L_D$ has the coefficient matrix $(N_{\L_D})_{\la_L,\la_R}
=\|W(g)\|\,\d_{\la_L\la_R}$. ($\|W(g)\|$ is the order of the Weyl group
of $g$.)

	If in addition the lattice $\L$ is both  even and self-dual, its
associated partition function $Z_\L$ is modular invariant. This is the key
property we are looking for.  The RTW lattice method suggests  we generate
all these functions (in the Appendix we describe how we did so). The resulting
invariants are not necessarily linearly
independent, nor are they immediately consistent with conditions (P2) or (P3).
But it is always possible to deduce from them a linear independent set and to
construct linear combinations

$$Z(z_Lz_R|\u)=\sum_\L A_\L \, Z_{\L}(g,k)(z_Lz_R|\u) \eqno(2.7)$$

\noindent  that satisfy both (P2) and (P3) (precisely how we do this is
discussed at the end of the Appendix). Such sums will certainly be  physical
invariants provided each $\L$ included in the sum
is an even self-dual gluing of $\L\*{g,k}$.  The
set of the partition functions $Z_{\L}(g,k)$ corresponding to all even
self-dual gluings $\L$ of $\L\!\*{g,k}$ is non-empty (containing at least the
diagonal invariant) and finite (its content being limited by the level and the
rank of the algebra).  As shown in \COMM{} this set is complete: it spans
the {\it Weyl-folded commutant\/} of $(g,\,k)$ -- \ie the complex space of all
partition functions (1.1) invariant under the modular group. Thus the lattice
method will allow us to compute all possible, and hence all physical,
invariants for a given $(g,\,k)$.

      Incidentally, this RTW lattice method has been generalized in \HET{} to
heterotic modular invariants, where it is also proven to be complete. It can
also be used to find modular invariants for the coset models.

\bigskip \bigskip
	      \centerline{{\bf III. INVARIANTS OF THE RANK-TWO ALGEBRAS }}
\bigskip\nobreak
	The above formalism can be applied in principle to any Lie algebra
$g$ at any level $k$.  However, since the dimension of the corresponding
lattices increases with the rank $r$ of the algebra and the members of the set
$P_+(g,k)$ grow rapidly in number with $r$ and level $k$, we will limit
ourselves in this work to the four rank-two algebras, $A_2$, $A_1+A_1$, $G_2$,
and $C_2$, and to levels approximately less than 30.  Following the approach
outlined above, we have calculated the partition functions for all of the even
self-dual gluings of
$(A_2\*{(k+3)};\, A_2\*{(k+3)})$ and
$(A_1\*{(k+2)}\oplus A_1\*{(k+2)};\,A_1\*{(k+2)}\oplus A_1\*{(k+2)})$
for various
values $k=1,2,\ldots\,$. From these results, we deduced the corresponding
invariants for $G_2$ and $C_2$. In each case, a set of linear independent
invariants was obtained and finally, by linear combinations, the physical
invariants satisfying conditions (P2) and (P3) were determined.

	In the algebra $g=A_2$, the two colabels are equal to 1, and so
the dual Coxeter number is $\h=3$ and the height $K=k+3$ at level $k$.  Let
$\beta_1$ and $\beta_2$ stand for the fundamental weights of $A_2$, and
$\rho=\beta_1+\beta_2$, their sum. The coroot ($=$ root) lattice is also
denoted by $A_2$, and its dual by $A_2\* *$. As usual, any
$\la=m\beta_1+n\beta_2
\in A_2\* *$ is identified through its Dynkin labels as $(m,n)$.

  The set of all possible horizontal highest weights corresponding to level $k$
is given by $$P_{+}(A_2,k)=
\{m\beta_1+n\beta_2\,|\,m,n\in\Z,\sp 0\le m,n,\sp m+n\le k\}.\eqno(3.1a)$$
\noindent
The trivial representation (the vacuum) at level $k$ defined
by the highest weight $\la=0$ is associated with the character
$\chi_{\rho}\*k=\chi_{11}\*k$, so (P3) reads $N_{11,11}=1$.
We will also refer to the set of highest weights
$$P(A_2,k)=P_+(A_2,k)+\rho=
\{m\beta_1+n\beta_2\,|\,m,n\in\Z,\sp 0< m,n,\sp m+n<k+3\}.\eqno(3.1b)$$

  The first step in our procedure is to find all self-dual gluings of
$(A_2\*{(K)};\,A_2\*{(K)})$, a step which we will describe in some detail in
the Appendix.  Next, to convert a lattice partition function (2.4) to the form
(1.1), it suffices to write $\c_\la$ for arbitrary $\la\in A_2\* *$ in terms of
$\c_{\la'}$ for $\la'\in P(A_2,k)$. This is not difficult once the action of
the six elements of $W(A_2)$ on $A_2\* *$ is written out explicitly.  For
example, $\c_{m,n}=0$ {\it iff\/}  $m\equiv 0$ (mod $K$), or $n\equiv 0$ (mod
$K$), or $m+n\equiv 0$ (mod $K$).

Finally, under the outer automorphism $h$ of the $A_2$ algebra, the character
transforms as
$$\c_{h(m,n)}=\c_{n,m}.\eqno(3.2a)$$
If $Z$ is any invariant with coefficient matrix $N$, we can construct
its {\it conjugation}, $Z\* c$, by
$$N\* c_{\la_L\la_R}=N_{\la_L,h(\la_R)}.\eqno(3.2b)$$
The conjugation of any physical invariant is also a physical invariant. We
can similarly define two other invariants, ${}\* cZ$ and ${}\* cZ\* c$,
respectively  by
${}\* cN_{\la_L\la_R}= N_{h(\la_L),\la_R}$ and
${}\* cN_{\la_L\la_R}\* c=N_{h(\la_L),h(\la_R)}$, but in this algebra one
always has ${}\* cZ=Z\* c$ and ${}\* cZ\* c=Z$.

       The fact that $G_2$ contains $A_2$ as a subalgebra of equal rank gives
us a many-to-one mapping from the $A_2$ invariants to the $G_2$ invariants
(this was also discussed in \WA).  In particular, let $\tilde \alpha_1$
($\tilde \alpha_2$) denote the long (short) simple root of $G_2$; then the
colabels are $a_1\*\vee=2$ and $a_2\*\vee=1$, and Kac-Peterson formula
$(2.2a$) tells us
	$${\tilde \c}_{m,n}=\c_{m,m+n}-\c_{m+n,m},\eqno(3.3)$$
where the character ${\tilde \c}$
belongs to $G_{2}\*{(1)}$ at level $k$, and the characters $\c$ belong to
$A_{2}\*{(1)}$ at level $k+1$. Given any ($k+1$)-level $A_2$ invariant $Z$ in
(1.1), we can construct a $k$-level $G_2$ invariant $\tilde Z$ by applying
(3.3) to $Z-Z\* c-{}\* cZ+{}\* cZ\* c$. Unfortunately, although this mapping
preserves
(P1), it does not necessarily preserve either (P2) or (P3). Nevertheless, all
$G_2$ invariants can be obtained in this way from $A_2$ invariants. Thus we can
read off the $G_2$ Weyl-folded commutant of level $k$ from the $A_2$
Weyl-folded commutant of level $k+1$; the physical invariants are then
obtained in the usual way.

Let us now turn to the case $g=A_1+A_1$. The lattice $M_g$ here is the
orthogonal direct sum $A_1\oplus A_1=\Z\*{(2)}\oplus \Z\*{(2)}$.  The two
fundamental weights of this algebra are $\beta_1=(1/\sqrt{2},0)$ and
$\beta_2=(0,1/\sqrt{2})$. The vector $\rho$ is as usual $\beta_1+\beta_2$. The
integrable highest weight representations of the Kac-Moody algebra $\g=\hat
A_1+\hat A_1$ have two levels, $k_1$ and $k_2$, one for each $\hat A_1$ factor.
Define the set

	$$P_{+}(A_1+A_1,k_1,k_2)=\{m\beta_1+n\beta_2\,
|\,m,n\in\Z,\sp 0\le m\le k_1,\sp 0\le n\le k_2\}.\eqno(3.4)$$

{\noindent The} characters of this algebra, $\chi_{m,n}\*{k_1,k_2}$, are
labelled by $(m,n)$.  They are just products of the characters of $\hat
A_1\*{(1)}$
$$\c_{m,n}\*{k_1,k_2}= \c_{m}\*{k_1}\c_{n}\*{k_2}.\eqno(3.5)$$

When $k_1=k_2$ (the case we will be considering), there is a conjugation
that interchanges the two $A_1$ factors. It maps a weight
$(m,n)$ to $(n,m)$, as in (3.2$a$). Through this mapping, any invariant $Z$
can be associated with three other invariants, $Z\* c$, ${}\* cZ$, and
${}\* cZ\* c$, as
for $A_2$. However, unlike for $A_2$, these four invariants will
all be distinct in general.

Because $C_2$ contains $A_1+A_1$ as an equal rank subalgebra, the same trick
used for $G_2$ will equally work here. The colabels are $a_1\*\vee=a_2\*\vee
=1$. A formula analogous to (3.3) holds. It
defines a surjective map from the $(k+1,k+1)$-level $A_1+A_1$ Weyl-folded
commutant to the $k$-level  $C_2$ Weyl-folded commutant as before.

	There is a property already discussed in \COMM{} (also found in
\RTWb) that proves to be quite useful in practice.  Consider any
$\la_L-\rho,\,\la_R-\rho \in P_+(g,k)$ and $\L\*{g,k}$, the Lorentzian root
lattice of $g$ scaled up in the usual way.  We call a positive integer $L$ the
order of $(\la_L;\la_R)$ in $\L\*{g,k}$ when, for any integer $m$,
$(m\la_L;m\la_R) \in \L_k$ {\it iff\/} $L$ divides $m$.  For each  $\ell$
relatively prime to $L$, $(\ell\la_L;\,\ell\la_R)$ is related nontrivially to
$(\la_L;\,\la_R)$, that is,
$\tilde\epsilon(\ell\la_L)\tilde\epsilon(\ell\la_R)\ne 0$ and
$$N_{\la_L\la_R}=\tilde\epsilon(\ell\la_L)\tilde\epsilon(\ell\la_R)
N_{[\ell\la_L][\ell\la_R]}\eqno(3.6)$$
($\tilde \epsilon$ and $[\la]$ are defined in (2.3$a$, $b$)). See \COMM{} for
examples.

This property has two valuable consequences. All such
$([\ell\la_L];\,[\ell\la_R])$ form a family of essentially equivalent
representations, so that just one representative of each family need be stored.
Another practical implication of (3.6) is that if for  some $\ell$,
$\tilde\epsilon(\ell\la_L)\tilde\epsilon(\ell\la_R)=-1$, then
$N_{\la_L\la_R}=0$ for any {\it positive\/} invariant $N$.

Call the family $([\ell \la_L];\,[\ell \la_R])$ a {\it positive parity
family\/} if $\tilde \epsilon(\ell\la_L)\tilde\epsilon(\ell\la_R)=1$ for all
$\ell$ relatively prime to the order $L$. By the {\it positive parity
commutant\/} we mean the subspace of the Weyl-folded commutant consisting of
all invariants $Z$ with the property that $N_{\la_L\la_R}\ne 0$ only for
$(\la_L;\la_R)$ in positive parity families. The positive parity commutant
contains all positive invariants and so is the only part of the Weyl-folded
commutant we need to consider. As we shall see, it is generally significantly
smaller than the full Weyl-folded commutant.

\bigskip\bigskip\centerline{\bf IV. THE PHYSICAL INVARIANTS}
\bigskip\nobreak

Most of this section is devoted to listing all the physical invariants we
found.  Our results are summarized in Tables 1 and 2. We give there the
dimensions of the Weyl-folded and positive parity commutants, along with the
numbers of physical invariants for each algebra and level we considered. The
positive parity commutant was defined at the end of the previous section; it
necessarily contains all the physical invariants and was the subspace we
focused on. The tables show the smallness of its dimension $P$, especially for
$A_2$, $G_2$ and $C_2$, compared with the dimension $D$ of the Weyl-folded
commutant.
This strongly suggests that a key component of a classification proof for
these (and probably all) algebras is eq.(3.6). Indeed that is the case with
the $A_2$ classification proof in \GA.

It seems to be quite difficult to obtain a general formula for the dimension
$D$ of the Weyl-folded commutant (though much more is known \BI{} about a
related space, the {\it pre-Weyl-folded commutant}). Tables 1 and 2 suggest
the following formulae for $D$ at prime heights $K=k+\h$:
$$\eqalign{{\rm for}\sp A_2:&\sp D=\left\{ \matrix{(K+5)/3& {\rm when}\sp
K\equiv 1\sp ({\rm mod}\sp 3)\cr (K+1)/3& {\rm when}\sp K\equiv 2\sp ({\rm
mod}\sp 3)\cr}\right. ;\cr
{\rm for}\sp G_2:&\sp D=\left\{ \matrix{(K+5)/6& {\rm when}\sp
K\equiv 1\sp ({\rm mod}\sp 3)\cr (K+1)/6& {\rm when}\sp K\equiv 2\sp ({\rm
mod}\sp 3)\cr}\right. ;\cr
{\rm for}\sp A_1+A_1:&\sp D=\left\{ \matrix{(K+3)/2& {\rm when}\sp
K\equiv 1\sp ({\rm mod}\sp 4)\cr (K+1)/2& {\rm when}\sp K\equiv 3\sp ({\rm
mod}\sp 4)\cr}\right. ;\cr
{\rm for}\sp C_2:&\sp D=\left\{ \matrix{(K+3)/4& {\rm when}\sp
K\equiv 1\sp ({\rm mod}\sp 4)\cr (K+1)/4& {\rm when}\sp K\equiv 3\sp ({\rm
mod}\sp 4)\cr}\right. .\cr}$$
Of course all
these formulae only apply to {\it prime} $K$; formulae for composite
heights will be more complicated and are presently unknown (with one
exception noted below). The above formula for
$A_2$ is proven in \RTW. Ph.{} Ruelle \RUE{} has also obtained the formula
for $G_2$ given above. In addition, at the heights $K=p\* 2$ of $A_2$, where
$p\equiv 5$ (mod 12) is {\it prime}, E.{} Thiran \RUE{} has found the formula
$D=2(7q\* 2
-3q+1)$ where $q=(K+1)/6$. For $k=22$ (the only point of intersection of
Thiran's formula and our work), both this formula and Table 1 gives $D=10$.

One curiosity we obtained was that all invariants, physical or otherwise,
for all the algebras and levels we considered were {\it symmetric} -- \ie their
coefficient matrices satisfied $N_{\la\la'}= N_{\la'\la}$. This relation is not
true of all algebras (\eg $g=D_{4n}$), but whenever it does hold, an easy
calculation shows that the coefficient matrices of all invariants will commute
with each other, so it could be a valuable theoretical property when studying
the commutant. We have been able to prove, using the analysis in \RTW, that all
invariants of $A_2$ height $K$ will be symmetric whenever $K$ is a product of
distinct primes and is not divisible by 3 (a similar result will then
necessarily hold for these heights of $G_2$), but the case of general $K$
seems more difficult.

Letting $P\* k$ denote the set $P(A_2,k)$ in $(3.1b$) and dropping the
superscripts `$A_2,k$' from the characters, the known physical invariants of
$A_2$ become:

$$\eqalignno{
  {\cal A}_k\equi & \sum_{\la\in P\* k} |\c_\la|\*2;  &(4.1a)\cr
{\cal D}_k\equi & \sum_{(m,n)\in P\* k}\c_{m,n}\c_{\omega\*{k(m-n)}(m,n)}\*{*},
{\rm \sp for \sp}k\not\equiv 0 \sp{\rm (mod\sp 3) \sp and\sp}k\ge 4;&(4.1b)\cr
 {\cal D}_k\equi &{1\over3}\sum_{{(m,n)\in P\* k\atop m\equiv n\sp({\rm mod}\sp
   3)}}      |\c_{m,n}+\c_{\omega(m,n)}+\c_{\omega\*2(m,n)}|\*2
    {\rm \sp for \sp}k\equiv 0 \sp{\rm (mod\sp 3)};   &(4.1c)\cr
  {\cal E}_5\equi & |\c_{1,1}+\c_{3,3}|\*2+|\c_{1,3}+\c_{4,3}|\*2+|\c_{3,1}
    +\c_{3,4}|\*2&\cr & +|\c_{3,2}+\c_{1,6}|\*2+|\c_{4,1}+\c_{1,4}|\*2
    +|\c_{2,3} +\c_{6,1}|\*2;  &(4.1d)\cr
  {\cal E}_9\*{(1)}\equi &|\c_{1,1}+\c_{1,10}+\c_{10,1}+\c_{5,5}+\c_{5,2}
    +\c_{2,5}|\*2& \cr &+2|\c_{3,3}+\c_{3,6}+\c_{6,3}|\*2;&(4.1e)\cr
  {\cal E}_9\*{(2)}\equi &|\c_{1,1}+\c_{10,1}+\c_{1,10}|\*2+|\c_{3,3}+\c_{3,6}+
    \c_{6,3}|\*2+2|\c_{4,4}|\*2& \cr
    &+|\c_{1,4}+\c_{7,1}+\c_{4,7}|\*2+|\c_{4,1}+
    \c_{1,7}+\c_{7,4}|\*2+|\c_{5,5}+\c_{5,2}+\c_{2,5}|\*2&\cr &+(\c_{2,2}
    +\c_{2,8}+\c_{8,2})\c_{4,4}\* *+\c_{4,4}(\c_{2,2}\* *+\c_{2,8}\* *
    +\c_{8,2}\* *);&(4.1f)\cr
  {\cal E}_{21}\equi & |\c_{1,1}+\c_{5,5}+\c_{7,7}+\c_{11,11}+\c_{22,1}
    +\c_{1,22}&\cr &+\c_{14,5}+\c_{5,14}+\c_{11,2}+\c_{2,11}+\c_{10,7}
    +\c_{7,10}|\*2&\cr &+|\c_{16,7}+\c_{7,16}+\c_{16,1}+\c_{1,16}+\c_{11,8}
    +\c_{8,11} &\cr &
    +\c_{11,5}+\c_{5,11}+\c_{8,5}+\c_{5,8}+\c_{7,1}+\c_{1,7}|\*2;&(4.1g)\cr}$$
together with their conjugations $Z\* c$. The subscript in ${\cal A}_k$ etc.{}
denotes the level of the invariant.
Note that ${\cal D}_3={\cal D}_3\* c$, ${\cal D}_6={\cal D}_6\* c$,
${\cal E}_9\*{(1)}={\cal E}_9\*{(1)}{}\* c$, and ${\cal E}_{21}=
{\cal E}_{21}\* c$.
The invariant $(4.1b)$ was first found in \ALZ, while $(4.1c$) was found in
\BER. The results $(4.1d,e,g$) were found in \CR, and $(4.1f$) was found in
\MS.

This list was proven complete in \GA{} for all but finitely many levels. The
remaining levels are all covered by this program, so we now know eqs.(4.1)
give the complete list of $A_2$ physical invariants.

\medskip
Now letting $P\* k$ denote the set $\rho+P_+(G_2,k)$ and again
dropping superscripts from the characters, the known
$G_2$ physical invariants are the series
    $${\cal A}_k\equi  \sum_{\la\in P\* k} |\c_\la|\*2,  \eqno(4.2a) $$
two invariants due to conformal embeddings at $k=3$ and $k=4$ (found by \CR)
$$\eqalignno{
  {\cal E}_3\equi& |\c_{1,1}+\c_{2,2}|\*2+ 2|\c_{1,3}|\*2,&(4.2b)\cr
  {\cal E}_4\*{(1)}\equi & |\c_{1,1}+\c_{1,4}|\*2+|\c_{2,1}+\c_{1,5}|\*2
	 +2|\c_{2,2}|\*2,&(4.2c)\cr}$$
and another invariant at $k=4$ (found in \VE)
  $$\eqalignno{{\cal E}_4\*{(2)}\equi & |\c_{1,1}|\*2+|\c_{2,2}|\*2
    +|\c_{1,3}|\*2+|\c_{2,3}|\*2 +|\c_{1,4}|\*2\cr
    & +\c_{2,1}\c_{1,5}\* *+\c_{1,5}\c_{2,1}\* *+
     \c_{3,1}\c_{1,2}\* *+\c_{1,2}\c_{3,1}\* *.&(4.2d )\cr }$$
Eqs.{} (4.2) exhaust all known $G_2$ physical invariants. This list has
been proven complete for all {\it prime} heights $K=k+4$ satisfying
$K\equiv 5,7$ (mod 12) \RUE; of course our program verifies its completeness
for all other $k\le 31$.

\medskip
   For the algebra $A_1+A_1$, we have only investigated the case $k_1=k_2 =k=
K-2$. The physical invariants belonging to infinite series are all simple
current invariants \SY, and have been described in \FKSV.  We shall label
them using simple current notation. These invariants may be given in
terms of their coefficient matrices $N_{ij,i'j'}$.

    In the next few paragraphs, we mean by ${\cal A}_k$, ${\cal D}_k$, ${\cal
E}_{10}$, ${\cal E}_{16}$, and ${\cal E}_{28}$ the various $A_1$ physical
invariants \CIZ. Tensor products of their coefficient matrices are coefficient
matrices of $A_1+A_1$ physical invariants.  These tensor products give rise to
some of the series solutions. In all, there are 4 series when $k$ is odd and
12 series
when $k$ is even, and a number of exceptional invariants at various levels.

\smallskip For $k$ odd, the 4 series are:

\noindent (i)  the diagonal invariant $N_k(0)\equi
{\cal A}_k\otimes {\cal A}_k$;

\noindent (ii) the invariant $N_k(J_1J_2)$, for $k>1$, defined by
$$          N_k(J_1J_2)_{ij,i'j'}\equi\left\{ \matrix{
			 \d_{ii'}\d_{jj'} & {\rm if}\sp i\equiv j \sp
			({\rm mod}\sp 2)\cr
			 \d_{i,K-i'}\d_{j,K-j'} & {\rm otherwise}\cr}
			 \right.;  \eqno(4.3a)$$
\noindent as well as their conjugations $N_k(0)\* c$ and $N_k(J_1J_2)\* c$.

\smallskip For $k\equiv 2$ (mod 4), there are 12 series:

\noindent (i) the diagonal invariant $N_k(0)$;

\noindent (ii) the invariant $N_k(J_1)\equi{\cal D}_k\otimes {\cal A}_k$
		${\rm \sp for} \sp k>2$;

\noindent (iii) the invariant $N_k(J_2)\equi{\cal A}_k\otimes {\cal D}_k$
		${\rm \sp for} \sp k>2$;

\noindent (iv) the invariant $N_k(J_1J_2)$ defined by
$$        N_k(J_1J_2)_{ij,i'j'}\equi\left\{ \matrix{
			 0 & {\rm if}\sp i\not\equiv j \sp
				      ({\rm mod}\sp 2)\cr
	     \d_{ii'}\d_{jj'}+\d_{i,K-i'}\d_{j,K-j'} & {\rm otherwise}
				      \cr}\right.;\eqno(4.3b)$$

\noindent (v) the invariant $N_k(J_1;J_2)\equi{\cal D}_k\otimes {\cal D}_k$
		${\rm \sp for} \sp k>2$;

\noindent (vi) the invariant $N_k(J_1;J_1J_2)$, for $k>2$, defined by
$$        N_k(J_1;J_1J_2)_{ij,i'j'}\equi\left\{
	    \matrix{  0
	      & {\rm if}\sp i\not\equiv j\sp({\rm mod}\sp 2)\cr
		 \d_{ii'}\d_{jj'}+\d_{i,K-i'}\d_{j,K-j'}
	      &{\rm if}\sp i\equiv j\equiv 1\sp({\rm mod}\sp 2)\cr
		\d_{i,K-i'}\d_{jj'}+\d_{ii'}\d_{j,K-j'}
	      &{\rm if}\sp i\equiv j\equiv 0 \sp({\rm mod}\sp 2)\cr}
		  \right.; \eqno(4.3c)$$

\noindent together with their conjugations $N\* c$.
We find $N_2(J_1J_2)\* c=N_2(J_1J_2)$.

\smallskip For $k\equiv 0$ (mod 4), the 12 series are:

\noindent (i) the diagonal invariant $N_k(0)$;

\noindent (ii) the invariant $N_k(J_1)\equi{\cal D}_k\otimes {\cal A}_k$;

\noindent (iii) the invariant $N_k(J_2)\equi{\cal A}_k\otimes {\cal D}_k$;

\noindent (iv) the invariant $N_k(J_1J_2)$ defined by
$$        N_k(J_1J_2)_{ij,i'j'}\equi\left\{ \matrix{
			 0 & {\rm if}\sp i\not\equiv j\sp({\rm mod}\sp 2)\cr
	     \d_{ii'}\d_{jj'}+\d_{i,K-i'}\d_{j,K-j'} & {\rm otherwise}
			\cr}\right.; \eqno(4.3d)$$

\noindent (v) the invariant $N_k(J_1;J_2)\equi{\cal D}_k\otimes {\cal D}_k$;

\noindent (vi) the invariant $N_k(aut)$ defined by
$$        N_k(aut)_{ij,i'j'}\equi\left\{ \matrix{
	      \d_{ii'}\d_{jj'} & {\rm if}\sp i\equiv j\equiv 1 \sp
		      ({\rm mod}\sp 2)\cr
	      \d_{i,K-i'}\d_{j,j'} & {\rm if}\sp i\equiv 1,\sp j\equiv 0\sp
		      ({\rm mod}\sp 2)\cr
	      \d_{ii'}\d_{j,K-j'} & {\rm if}\sp i\equiv 0,\sp j\equiv 1\sp
		       ({\rm mod}\sp 2)\cr
	      \d_{i,K-i'}\d_{j,K-j'} & {\rm if}\sp i\equiv j\equiv 0\sp
		       ({\rm mod}\sp 2)\cr}\right.;\eqno(4.3e)$$

\noindent together with their conjugations $N\* c$.

Besides these series, we also obtain a number of $A_1+A_1$ exceptional
physical invariants. They occur at any level $k$ where there exists
an $A_1$ exceptional invariant; invariants of this type are
given by the tensor products ${\cal A}_{10}\otimes{\cal E}_{10}$,
${\cal D}_{10}\otimes{\cal E}_{10}$, ${\cal E}_{10}\otimes{\cal A}_{10}$,
${\cal E}_{10}\otimes{\cal D}_{10}$, ${\cal E}_{10}\otimes{\cal E}_{10}$,
${\cal A}_{16}\otimes{\cal E}_{16}$, ${\cal D}_{16}\otimes{\cal E}_{16}$,
${\cal E}_{16}\otimes{\cal A}_{16}$, ${\cal E}_{16}\otimes{\cal D}_{16}$
and ${\cal E}_{16}\otimes{\cal E}_{16}$, and their respective conjugations
$Z\* c$.  There exist further sporadic exceptional physical invariants not of
this type, which we denote by ${\cal E}''_k$ to distinguish them from the $A_1$
exceptionals. They are:

$$\eqalignno{
{\cal E}''_4\equi& |\c_{1,1}+\c_{1,5}+\c_{5,1}+\c_{5,5}|\*2+
  2|\c_{1,3}+\c_{5,3}|\*2+2|\c_{3,1}+\c_{3,5}|\*2+4|\c_{3,3}|\*2;&(4.3f)\cr
{\cal E}''_6\equi&|\c_{1,1}+\c_{3,5}+\c_{5,3}+\c_{7,7}|\*2+|\c_{3,3}+\c_{7,1}+
  \c_{1,7}+\c_{5,5}|\*2&\cr
    &+|\c_{2,4}+\c_{4,2}+\c_{6,4}+\c_{4,6}|\*2;&(4.3g)\cr
{\cal E}''_8\equi&|\c_{1,1}+\c_{1,9}+\c_{9,1}+\c_{9,9}|\*2+|\c_{3,3}+\c_{3,7}
  +\c_{7,3}+\c_{7,7}|\*2&\cr
  &+(\c_{1,3}+\c_{1,7}+\c_{9,3}+\c_{9,7}+\c_{3,1}+\c_{3,9}+\c_{7,1}+\c_{7,9})
  \c_{5,5}\* *&\cr
  &+\c_{5,5}(\c_{1,3}+\c_{1,7}+\c_{9,3}+\c_{9,7}+\c_{3,1}+\c_{3,9}+\c_{7,1}+
  \c_{7,9})\* *&\cr
  &+2|\c_{5,5}|\*2+|\c_{3,5}+\c_{7,5}+\c_{5,3}+\c_{5,7}|\*2+|\c_{1,5}+\c_{9,5}
  +\c_{5,1}+\c_{5,9}|\*2;&(4.3h)\cr
{\cal E}''_{10}{}\*{(1)}\equi&|\c_{1,1}+\c_{1,7}+\c_{5,5}+\c_{5,11}+\c_{7,1}
   +\c_{7,7}+\c_{11,5}+\c_{11,11}|\*2&\cr&
   +|\c_{1,11}+\c_{1,5}+\c_{7,11}+\c_{7,5}+
   \c_{5,7}+\c_{5,1}+\c_{11,1}+\c_{11,7}|\*2&\cr
   &+2|\c_{4,4}+\c_{4,8}+\c_{8,4}+\c_{8,8}|\*2;&(4.3i)\cr
{\cal E}''_{10}{}\*{(2)}\equi&|\c_{1,1}+\c_{1,7}+\c_{11,5}+\c_{11,11}|\*2+
   |\c_{5,5}+\c_{5,11}+\c_{7,1}+\c_{7,7}|\*2&\cr
   &+|\c_{1,5}+\c_{1,11}+\c_{11,1}+\c_{11,7}|\*2+|\c_{5,1}+\c_{5,7}+\c_{7,5}+
   \c_{7,11}|\*2&\cr
   &+|\c_{2,4}+\c_{2,8}+\c_{10,4}+\c_{10,8}|\*2+|\c_{4,4}+\c_{4,8}+\c_{8,4}+
   \c_{8,8}|\*2+2|\c_{6,4}+\c_{6,8}|\*2&\cr
  &+|\c_{3,1}+\c_{3,7}+\c_{9,5}+\c_{9,11}|\*2+|\c_{3,5}+\c_{3,11}+\c_{9,1}+
  \c_{9,7}|\*2;&(4.3j)\cr}$$
together with the three conjugations ${\cal E}''_{10}{}\*{(2)}{}\* c$, ${}\* c
{\cal E}''_{10}{}\*{(2)}$ and ${}\* c{\cal E}''_{10}{}\*{(2)}{}\* c$ of
$(4.3j$).

The invariant $(4.3g$) is due to a conformal embedding. The invariant
$(4.3f$) was found in \FKSV,
while $(4.3h$) was found in \VE. The invariant
$(4.3i$) can be understood as due to the
conformal embedding of $A_{1}+A_{1}$ level (10,10) into $C_{2}+C_{2}$ level
(1,1), applied to a
non-diagonal physical invariant of the latter.
The invariant $(4.3j$) can be written as
the product of the coefficient matrix of ${\cal A}_{10}
\otimes {\cal E}_{10}$ with $N_{10}(J_1J_2)$ (see $(4.3b$)). Note that $k=10$
has an incredibly large number of 27 physical invariants.

Eqs.{} (4.3) exhaust all the $A_1+A_1$ physical invariants for $k_1=k_2\le 21$.
The only other known ones for $k_1=k_2$ occur at $k=28$.

\medskip As before let $P\* k$ denote the set $\rho+P_+(C_2,k)$.
The $C_2$ physical invariants include two series of solutions, namely,
$$\eqalignno{
  {\cal A}_k\equi & \sum_{\la\in P\* k} |\c_\la|\*2,&(4.4a )\cr
  {\cal D}_k\equi & \sum_{{(m,n)\in P\* k \atop m\sp {\rm odd}}} |\c_{m,n}|\*2
       + \sum_{{(m,n)\in P\* k\atop m\sp {\rm even}}} \c_{m,n}\c_{m,K-m-n}\* *
		    {\rm \sp for} \sp k>1 {\rm\sp odd}, &(4.4b)\cr
  {\cal D}_k\equi & \sum_{{(m,n)\in P\* k\atop m \sp {\rm odd}}}
      \c_{m,n}\c_{m,n}\*{*}+\c_{m,n}\c_{m,K-m-n}\*{*}
      {\rm \sp for} \sp k {\rm\sp even}, &(4.4c)\cr     }$$
and four exceptional solutions:
$$\eqalignno{
{\cal E}_3\equi& |\c_{1,1}+\c_{3,2}|\*2+ |\c_{1,4}+\c_{3,1}|\*2
		   +2|\c_{2,2}|\*2;&(4.4d )\cr
{\cal E}_7\equi & |\c_{1,1}+\c_{1,6}+\c_{3,3}+\c_{7,2}|\*2&\cr
 &+|\c_{1,3}+\c_{1,8}+\c_{3,4}+\c_{7,1}|\*2+2|\c_{4,2}+\c_{4,4}|\*2;&(4.4e )\cr
  {\cal E}_{8}\equi & |\c_{1,1}+\c_{1,9}|\*2+|\c_{3,3}+\c_{3,5}|\*2+|\c_{5,5}
  +\c_{5,1}|\*2+|\c_{1,7}+\c_{1,3}|\*2+|\c_{5,4}+\c_{5,2}|\*2&\cr
&+|\c_{9,1}|\*2+\c_{9,1}(\c_{1,2}+\c_{1,8})\* *+(\c_{1,2}+\c_{1,8})\c_{9,1}\* *
  +|\c_{5,3}|2+\c_{5,3}(\c_{3,6}+\c_{3,2})\* *&\cr
&+(\c_{3,6}+\c_{3,2})\c_{5,3}\* *+|\c_{1,5}|\*2+\c_{1,5}(\c_{3,7}+\c_{3,1})\* *
  +(\c_{3,7}+\c_{3,1})\c_{1,5}\* *+|\c_{3,4}|\*2&\cr
  &+\c_{3,4}(\c_{7,1}+\c_{7,3})\* *+(\c_{7,1}+\c_{7,3})\c_{3,4}\* *
   +|\c_{7,2}|\*2&\cr
  &+\c_{7,2}(\c_{1,4}+\c_{1,6})\* *+(\c_{1,4}+\c_{1,6})\c_{7,2}\* *;&(4.4f )\cr
{\cal E}_{12}\equi & |\c_{1,1}+\c_{1,13}+\c_{3,4}+\c_{3,8}
    +2\c_{5,5}+\c_{7,1}+\c_{7,7}+\c_{9,2}+\c_{9,4}|\*2.&(4.4g )\cr
}$$
Eqs.(4.4$b,c$) are found in \BER, $(4.4d,e,g$) are conformal embeddings
found in \CR, and $(4.4f$) is found in \VE. Eqs.{} (4.4) exhaust all known
$C_2$ physical invariants, and is the complete list for $k\le 26$.

\bigskip\bigskip\centerline{\bf V. CONCLUSION}
\bigskip\nobreak
In this paper we present a numerical search for physical invariants of the
rank-two algebras $A_2$, $A_1+A_1$, $G_2$, and $C_2$ at levels approximately
less than 30.  Within these limits, made necessary by the modest capacity of
our personal computer, we have shown first that the self-dual lattice method is
a useful and practical tool; in particular, it uncovers in these algebras and
at these levels, {\it all} physical invariants, those belonging to the series
as well
as those lying outside the series.  Secondly, we have shown that it can
fruitfully complement a purely analytical approach.  For example, it can
identify the physical invariants at those levels of $A_2$ which eluded
theoretical arguments.  It also can uncover regularities and suggest approaches
useful for these theoretical arguments. Furthermore, not only does it
determine the
dimensions of various commutants at specific levels but it also can
suggest general formulas applicable to many other, and in some cases all,
levels. It is clear
that the analytical method and the present numerical method, especially when
the restrictions on the ranks and levels of the algebras are relaxed, can
successfully complement each other in working towards a solution of the
classification of the conformal field theories.

\bigskip\bigskip
\centerline{\bf ACKNOWLEDGMENTS}\bigskip
\nobreak
This work was supported in part by the Natural Sciences and Engineering
Research Council of Canada. T.G.{} would like to thank the hospitality
of the Carleton mathematics department, as well as Patrick Roberts and
Philippe Ruelle for several valuable conversations.

\vfill\eject
\centerline {Table 1. Dimensions of the commutants and numbers of}
\centerline{physical invariants for $A_2$ and $G_2$}
\medskip
$$\vbox{\tabskip=0pt\offinterlineskip
  \def\tablerule{\noalign{\hrule}}
  \halign to 5.75in{
    \strut#&\vrule#\tabskip=0em plus2em &    
    \hfil#&\vrule#&\hfil#&\vrule#&    
    \hfil#&\vrule#&\hfil# &\vrule#&    
    \hfil#&\vrule#&\hfil# &\vrule#&    
    \hfil#&\vrule#\tabskip=0pt\cr\tablerule     &    
&&& \multispan5\hfil $A_2$\hfil &&\multispan5\hfil $G_2$\hfil &\cr\tablerule
&&\omit\hidewidth k\hidewidth&&
 \omit\hidewidth D\hidewidth&&  \omit\hidewidth P\hidewidth&&
\omit\hidewidth N\hidewidth&&  \omit\hidewidth D\hidewidth&&
 \omit\hidewidth P\hidewidth&&  \omit\hidewidth N\hidewidth& \cr\tablerule
&& 1 && 2 && 2 &&2&& 1  && 1 &&1& \cr
&& 2 && 2 && 2 &&2&& 1  && 1 &&1& \cr
&& 3 && 4 && 4 &&3&& 2  && 2 &&2& \cr
&& 4 && 4 && 4 &&4&&  2 && 3 &&3& \cr
&& 5 && 6 && 6 &&6&&  1 && 1 &&1& \cr
&& 6 && 3 && 3 &&3&&  3 && 1 &&1& \cr
&& 7 && 6 && 4 &&4&&  2 && 1 &&1& \cr
&& 8 && 4 && 4 &&4&&  4 && 1 &&1& \cr
&& 9 && 10 && 8 &&7&& 3  && 1 &&1& \cr
&& 10 && 6 && 4 &&4&& 4 && 1 &&1& \cr
&& 11 && 8 && 4 &&4&& 4  && 1 &&1& \cr
&& 12 && 9 && 4 &&4&& 6  && 1 &&1& \cr
&& 13 && 12 && 4 &&4&& 3 && 1 &&1& \cr
&& 14 && 6 && 4 &&4&&  5 && 1 &&1& \cr
&& 15 && 11 && 4 &&4&& 4  && 1 &&1& \cr
&& 16 && 8 && 4 &&4&&  10 && 1 &&1& \cr
&& 17 && 21 && 7 &&4&&  6 && 1 &&1& \cr
&& 18 && 12 && 5 &&4&&  6 && 1 &&1& \cr
&& 19 && 12 && 4 &&4&&  4 && 1 &&1& \cr
&& 20 && 8 && 4 &&4&&  12 && 1 &&1& \cr
&& 21 && 26 && 9 &&5&&  5 && 1 &&1& \cr
&& 22 && 10 && 4 &&4&&  7 && 1 &&1& \cr
&& 23 && 14 && 4 &&4&&  5 && 1 &&1& \cr
&& 24 && 12 && 4 &&4&&  15 && 1 &&1& \cr
&& 25 && 26 && 4 &&4&&  5 && 1 &&1& \cr
&& 26 && 10 && 4 &&4&&  21 && 1 &&1& \cr
&& 27 && 28 && 4 &&4&&  6 && 1 &&1& \cr
&& 28 && 12 && 4 &&4&&  13 && 1 &&1& \cr
&& 29 && 24 && 4 &&4&&  9 && 1 &&1& \cr
&& 30 && 17 && 4 &&4&&  9 && 1 &&1& \cr
&& 31 && 18 && 4 &&4&&  11 && 1 &&1& \cr
&& 32 && 22 && 4 &&4&&   &&  &&& \cr
\tablerule\noalign{\smallskip}
 }} $$
\centerline{$k=$ level; $D=$ dimension of the Weyl-folded commutant;}
\centerline{ $P=$ dimension of the positive parity commutant; and
$N=$ number of physical invariants.}\vfill\eject

\centerline {Table 2. Dimensions of the commutants and numbers of}
\centerline{physical invariants for $A_1+A_1$ and $C_2$}
\medskip
$$\vbox{\tabskip=0pt\offinterlineskip
  \def\tablerule{\noalign{\hrule}}
  \halign to 5.75in{
    \strut#&\vrule#\tabskip=0em plus1em &    
    \hfil#&\vrule#&\hfil#&\vrule#&    
    \hfil#&\vrule#&\hfil# &\vrule#&    
    \hfil#&\vrule#&\hfil# &\vrule#&    
    \hfil#&\vrule#\tabskip=0pt\cr\tablerule     &    
&&&\multispan5\hfil $A_1+A_1$\hfil &&\multispan5\hfil $C_2$\hfil &\cr\tablerule
&&\omit\hidewidth k\hidewidth&&
 \omit\hidewidth D\hidewidth&&  \omit\hidewidth P\hidewidth&&
 \omit\hidewidth N\hidewidth&&  \omit\hidewidth D\hidewidth&&
 \omit\hidewidth P\hidewidth&&  \omit\hidewidth N\hidewidth& \cr\tablerule
&& 1 && 2 && 2 &&2&&  1  && 1 &&1& \cr
&& 2 && 3 && 3 &&3&&  2  && 2 &&2& \cr
&& 3 && 4 && 4 &&4&&  3  && 3 &&3& \cr
&& 4 && 9 && 9 &&13&& 2 && 2 &&2& \cr
&& 5 && 4 && 4 &&4&&  3 && 2 &&2& \cr
&& 6 && 10 && 10 &&13&& 3 && 2 &&2& \cr
&& 7 && 6 && 4 &&4&&  5 && 3 &&3& \cr
&& 8 && 13 && 11 &&13&& 3 && 3 &&3& \cr
&& 9 && 6 && 4 &&4&& 9&& 4 &&2& \cr
&& 10 && 27 && 25 &&27&& 4 && 2 &&2& \cr
&& 11 && 8 && 4 &&4&&   6  && 2 &&2& \cr
&& 12 && 17 && 10 &&12&& 9  && 3 &&3& \cr
&& 13 && 22 && 4 &&4&&   7 && 2 &&2& \cr
&& 14 && 20 && 11 &&12&& 5 && 2 &&2& \cr
&& 15 && 10 && 4 &&4&& 10  && 2 &&2& \cr
&& 16 && 33 && 19 && 22&&  5 && 2 &&2& \cr
&& 17 && 10 && 4 &&4&&   14 && 2 &&2& \cr
&& 18 && 44 && 18 &&12&& 11 && 2 &&2& \cr
&& 19 && 26 && 4 &&4&&   9 && 2 &&2& \cr
&& 20 && 25 && 10 &&12&& 6 &&2  &&2& \cr
&& 21 && 12 && 4 &&4&&   21 &&2  &&2& \cr
&& 22 &&  &&  &&&&        9 && 2 &&2& \cr
&& 23 &&  &&  &&&&       11 &&2  &&2& \cr
&& 24 &&  &&  &&&&       12 &&2 &&2& \cr
&& 25 &&  &&  &&&&       17 && 2 &&2& \cr
&& 26 &&  &&  &&&&       8 &&2  &&2& \cr
\tablerule\noalign{\smallskip}
 }} $$
\centerline{$k=$ level; $D=$ dimension of the Weyl-folded commutant;}
\centerline{ $P=$ dimension of the positive parity commutant; and
$N=$ number of physical invariants.}\vfill\eject

\bigskip\bigskip
\centerline {\bf APPENDIX}
\bigskip

We describe an algorithm for finding all even self-dual gluings of
$\Omega_k=(A_2\*{(k+3)};A_2\*{(k+3)})$, where $k$ is as usual the level.
The method can be generalized to other algebras.

This algorithm involves  the gluing construction of lattices \Csgl.
One of the powers of gluing is the ease with
which \eg self-duality can be verified.
A lattice is self-dual {\it iff} it is both integral and its determinant equals
one.  A gluing $\L=\L_0[g_1,\ldots,g_n]$ is
integral {\it iff} its base lattice
$\L_0$ is integral, its glue generators $g_i$ all lie in $\L_0\* *$, and the
dot products $g_i\cdot g_j$ are all integers. The following fact \Gal{} allows
for straightforward computation of the determinant $|\L|$ of a gluing:
$$|\L|=|\L_0|/\|G\|\* 2,$$
where $\|G\|$ is the order of the {\it glue group} $G=\L/\L_0$ (\ie the
additive group spanned (mod $\L_0$) by the glue generators $g_i$), or
equivalently the number of different glue classes $x+\L_0$ in $\L$.

Let $K=k+3$ and $\ell=3K$. Let $\alpha_1$ and $\beta_1$ be the
first simple root and its corresponding fundamental weight of $A_2$.  By
$(a,b)$ we mean the two-dimensional (Euclidean) vector
$(a,b)=a\beta_1/\sqrt{K} +b\alpha_1/\sqrt{K}\in A_2\*{(K)*}$ and, similarly,
$(a,b;\,c,d)$ the four-dimensional (Lorentzian) vector
$(a\beta_1/\sqrt{K} +b\alpha_1/\sqrt{K}; \,a\beta_1/\sqrt{K}
+b\alpha_1/\sqrt{K}) \in\Omega_k\* *$.
It can be shown that any vector in $A_2\*{(K)*}$
equals $(a,b)$ for some integers $a,b$; similarly for $\Omega_k\* *$ and
$(a,b;\,c,d)$. Any
even self-dual gluing $\L$ of  $\Omega_k$ has a maximal left-hand side
sublattice $\L_L$ defined by the  set $\L_L=\{x_L|(x_L;0)\in\L\}$ and a maximal
right-hand side
sublattice $\L_R$ defined by the  set $\L_R=\{x_R|(0;x_R)\in\L\}$. Both of
these lattices are {\it even} gluings of $A_2\*{(K)}$ and necessarily
have equal determinants: $|\L_L|=|\L_R|$. We will construct all $\L$ by
first constructing all possible $\L_L$ and $\L_R$.

The first step of the algorithm is to generate all {\it even} gluings of
$A_2\*{(K)}$. Any gluing of $A_2\*{(K)}$ can be generated by two independent
glue vectors $(a,b)$ and $(c,d)$, and, without loss of generality, these
vectors may be taken to belong respectively to certain sets $S_1$ and
$S_2$ to be described as follows.

The first set, $S_1$, consists of $(\ell,0)$ and all the vectors $(a,b)$
such that $0<a<\ell$,  $0\le b<K$, $a$ divides $\ell$, and
$a\*2/3+ab+b\*2\equiv 0$ (mod $K$).
The second set, $S_2$, consists of all vectors $(0,d)$ such that
$0<d\le K$, $d$ divides $K$, and $d\*2\equiv 0$ (mod $K$).

The possible lattices $\L_L$ and $\L_R$ are then given by all the triples
$a,\,b,\,d$ satisfying the following conditions: $(a,b)\in S_1$, $(0,d)\in
S_2$, $b<d$, $\ell b$ is divisible by $ad$, and $(a+2b)d\equiv 0 $ (mod $K$).
The three congruences (mod $K$) correspond to the conditions that the glues
$(a,b)$ and $(0,d)$ have even norms and integral dot products. The remaining
conditions are designed to force a unique choice of glue generators.
There is an obvious one-to-one correspondence between a triple $a,\,b,\,d$ and
an even gluing of $A_2\*{(K)}$, given by
$$a,\,b,\,d\rightarrow A_2\*{(K)}[(a,b),(0,d)]\equi \bigcup_{i,j\in \Z}\{
A_2\*{(K)}+i(a,b)+j(0,d)\}=\bigcup_{i=1}\*{\ell/a}\bigcup_{j=1}\*{K/d}\{
A_2\*{(K)}+i(a,b)+j(0,d)\}.$$
  It turns out that $(a,b)$ and $(0,d)$ form a basis for such a lattice.  We
are actually generating more lattices here than we need later on: several
lattices may lead to the same lattice sum (2.4) because of the
``Weyl-folding'' (summing over $W(g)$) involved in this calculation;
Weyl-equivalent gluings may be eliminated at this point.

The next step in the procedure is to choose the base lattice $(\L_L;\L_R)$
satisfying the determinantal condition $|\L_L|=|\L_R|$. This is equivalent to
choosing triples $a,\,b,\,d$ (associated with $\L_L$) and $a',\,b',\,d'$
(associated with $\L_R$) satisfying $ad=a'd'$.

To construct the four-dimensional {\it even self-dual} gluings $\L$,
we define first the integers $L_1=\ell/a$, $L_2=K/d$, $L_1'=\ell/a'$, and
$L_2'=K/d'$, and then the vectors
$$\eqalign{
	h_1&=(2L_1,\,-L_1\,)\cr
	h_2&=(-3L_2-2L_1b/d,\, 2L_2+L_1b/d\,)\cr
	h_1'&=(2L_1',\,-L_1'\,)\cr
	h_2'&=(-3L_2'-2L_1'b'/d',\, 2L_2'+L_1'b'/d'\,)\cr}$$
These satisfy $h_i\cdot(a,b)=h_i'\cdot(a',b')=\d_{i1}$ and $h_i\cdot(0,d)=
h_i'\cdot(0,d')=\d_{i2}$, so $h_1$ and $h_2$ ($h_1'$ and $h_2'$) form a
basis for $\L_L\* *$ ($\L_R\* *$).
Next we find all integers $w,\,x,\,y,\,z$, where $0\le w,\,y<a'$ and
$0\le x,\,z<d'$, such that the
two Lorentzian four-dimensional vectors, $(h_1;\,w h_1'+xh_2'\,)$ and
$(h_2;\,y h_1'+zh_2'\,)$, have even norms and integral dot products.
To each such choice of $w,\,x,\,y,\,z$ corresponds the gluing

$$\eqalign{
\L&=(\L_L;\,\L_R\,)\,[\,(h_1;\,w h_1'+xh_2'\,),\,(h_2;\,y h_1'+zh_2'\,)\,]\cr
  &=(A_2\*{(K)};\,A_2\*{(K)}\,)\,[\,(00;\,a',b'),\,(00;\,0,d'\,),\,
    (h_1;\,w h_1'+xh_2'\,),\,(h_2;\,y h_1'+zh_2'\,)\,]\cr
  &=\bigcup_{i=1}\*{L_1'}\bigcup_{j=1}\*{L_2'}\bigcup_{l=1}\*a\bigcup_{m=1}\*d
    \bigl(\Omega_k+i(00;\,a',b')+j(00;\,0,d')+l
    (h_1;\,w h_1'+xh_2')+m(h_2;\,y h_1'+zh_2')\bigr).\cr}$$

\noindent By construction, the resulting lattice is even and also
necessarily self-dual: the number of different glue classes in it is given
by $adL_1'L_2'=3K\*2$. All even self-dual gluings $\L$ of $\Omega_k$
correspond to some choice of $a,\,b,\,c,\,a',\,b',\,d'$ and $w,\,x,\,y,\,z$.

We will conclude with brief comments on how we find all
invariants satisfying (P2) and (P3). Using Gaussian elimination on the
set $\{Z_\L\}$ of lattice partition functions, we construct a basis for
the Weyl-folded commutant, as well as a basis $\{Z\*{(1)},\ldots,Z\*{(P)}\}$
for a subspace (the positive parity commutant defined at the
end of Sec.{} III) which necessarily contains all physical invariants.
We are interested in the linear combinations $Z=\sum A_i Z\*{(i)}$. Even
when the dimension $P$ of the commutant is large, it is trivial to
find the conditions on the $A_i$ equivalent to the demand that the
coefficients $N_{\la_L\la_R}$ of $Z$ be integers: \eg in almost all cases
we have considered, this condition is simply that each $A_i\in\Z$. In our
notation (P3) becomes the condition that $N_{\rho\rho}=1$; for our
bases, (P3) is always equivalent to the choice $A_1=1$. Using the remaining
positivity conditions $N_{\la_L\la_R}\ge 0$, we then find upper and lower
bounds for the remaining $A_i$ (the most common ones are $0\le A_i\le 1$
or $-1\le A_i\le 0$). We then use the computer to run through the
possible values of $A_i$, checking for positivity of all coefficients.
As we see in the tables, most dimensions $P$ are small, but even for
large $P$ this final computer search is fast.

\bigskip\bigskip
\centerline{{\bf REFERENCES}} \bigskip
\frenchspacing
\i{\CA} J.~L. Cardy, Nucl. Phys {\bf B270}, 186 (1986).
\item{\GW} D. Gepner and E. Witten, Nucl. Phys. {\bf B278}, 493 (1986).
\i{\FKSV} J. Fuchs, A. Klemm, M. Schmidt and D. Verstegen, Int. J. Mod. Phys.
A {\bf 7}, 2245 (1992).
\i{\KP} V. Kac and D. Peterson, Adv. Math. {\bf 53}, 125 (1984).
\item{\CIZ} A. Cappelli, C. Itzykson and  J.-B. Zuber,
Nucl. Phys. {\bf B280 [FS18]}, 445 (1987).
\item{\MS} G. Moore and N. Seiberg, Nucl. Phys. {\bf B313}, 16 (1989).
\i{\SY} A.~N. Schellekens and S. Yankielowicz, Nucl. Phys. {\bf B327}, 673
(1989);
\i{} A.~N. Schellekens, Phys. Lett. B {\bf 244}, 255 (1990);
\i{} B. Gato-Rivera and A.~N. Schellekens, Commun. Math. Phys. {\bf 145}, 85
(1992).
\i{\KA} A. Kato,  Mod. Phys. Lett. A{\bf 2}, 585 (1987);
\item{} A. Cappelli, C. Itzykson and  J.-B. Zuber,
Commun. Math. Phys. {\bf 113}, 1 (1987);
\item{} D. Gepner and Z. Qiu, Nucl. Phys. {\bf B285}, 423 (1987).
\i{\Kas} D. Kastor, Nucl. Phys. {\bf B280 [FS18]}, 304 (1987).
\i{\GA} T. Gannon, ``The classification of affine SU(3) modular invariant
partition functions'', (Carleton preprint, 1992).
\item{\ITZ} C. Itzykson, Nucl. Phys. (Proc. Suppl.) {\bf 5B}, 150
(1988);
\item{} P. Degiovanni, Commun. Math. Phys. {\bf 127}, 71 (1990).
\item{\COMM} T. Gannon, ``WZW commutants, lattices, and level-one
partition functions'', Nucl. Phys. {\bf B} (to appear).
\i{\VE} D. Verstegen, Nucl. Phys. {\bf B346}, 349 (1990).
\item{\WA} N.~P. Warner, Commun. Math. Phys. {\bf 130}, 205 (1990).
\item{\RT} P. Roberts, Phys. Lett. {\bf B244}, 429 (1990);
\item{} P. Roberts and H. Terao, Int. J. Mod. Phys. A {\bf 7}, 2207 (1992).
\item{\HET} T. Gannon, ``Partition functions for heterotic WZW
 conformal field theories'', Nucl. Phys. {\bf B} (to appear).
\item{\KAC} V.~G. Kac, {\it Infinite Dimensional Lie Algebras}, 3rd ed.,
(Cambridge University Press, Cambridge, 1990).
\item{\KMPS} S. Kass, R.~V. Moody, J. Patera and R. Slansky, {\it Affine Lie
Algebras, Weight Multiplicities, and Branching Rules} Vol.1 (University
of California Press, Berkeley, 1990).
\item{\CS} J.~H. Conway and N.~J.~A. Sloane, {\it Sphere packings,
Lattices and Groups}, (Springer-Verlag, New York, 1988).
\i{\LSW} W. Lerche, A.~N. Schellekens and N.~P. Warner, Phys. Reports
{\bf 177}, 1 (1989).
\item{\RTWb} Ph. Ruelle, E. Thiran and J. Weyers,  ``Implications of
an arithmetical symmetry of the commutant for modular invariants'', (DIAS
preprint STP-92-26, 1992).
\i{\BI} M. Bauer and C. Itzykson, Commun. Math. Phys. {\bf 127}, 617 (1990);
\i{} Ph. Ruelle, Commun. Math. Phys. {\bf 133}, 181 (1990);
\i{} M. Bauer, ``Aspects de l'invariance conforme'', Ph.D. thesis, Saclay.
\item{\RTW} Ph. Ruelle, E. Thiran and J. Weyers,
Comm. Math. Phys. {\bf 133}, 305 (1990).
\i{\RUE} (Ph. Ruelle, private communication);
\i{} Ph. Ruelle, Ph.D. thesis, Louvain-la-Neuve, September 1990.
\i{\ALZ} D. Altsch\"uler, J. Lacki and Ph. Zaugg, Phys. Lett. B {\bf 205},
281 (1988).
\i{\BER} D. Bernard, Nucl. Phys. {\bf B288}, 628 (1987).
\item{\CR} P. Christe and F. Ravanani, Int. J. Mod. Phys. A {\bf
4}, 897 (1989).
\i{\Gal} T. Gannon and C.~S. Lam, Rev. Math. Phys. {\bf 3}, 331 (1991).

\end